\documentclass[prb,twocolumn,showpacs,amsmath,amssymb,superscriptaddress]{revtex4}
\usepackage{graphicx}
\usepackage{dcolumn}
\usepackage{bm}
\usepackage{epsfig}

\begin{document}

%\preprint{APS/123-QED}

\title{Persistent dynamics in the $S=1/2$ quasi-one-dimensional chain
compound
Rb$_{4}$Cu(MoO$_{4}$)$_{3}$ probed with muon-spin relaxation}

\author{T. Lancaster}
\affiliation{University of Durham, Centre for Materials Physics, South Road, 
Durham, DH1 3LE, UK}
\author{P.J. Baker}
\author{F.L. Pratt}
\affiliation{ISIS Facility, Rutherford Appleton Laboratory, Chilton, 
Oxfordshire OX11 0QX, UK}
%\author{J.S. M\"oller}
\author{S.J. Blundell}
\author{W. Hayes}
\author{D. Prabhakaran}
\affiliation{Oxford University Department of Physics, Clarendon Laboratory, 
Parks Road, Oxford, OX1 3PU, UK}

\date{\today}

\begin{abstract}
We report the results of muon-spin relaxation measurements on the 
low-dimensional antiferromagnet Rb$_{4}$Cu(MoO$_{4}$)$_{3}$. No long-range magnetic
order is observed down to 50~mK implying a ratio $T_{\mathrm{N}}/J<0.005$ (where
$J$ is the principal exchange strength along the spin chains) and an effective ratio of interchain to 
intrachain exchange of $|J_{\perp}/J|<2 \times 10^{-3}$, making the material
an excellent realization of a one-dimensional quantum Heisenberg antiferromagnet. 
We probe the persistent spin excitations at low temperatures and find that ballistic
spin transport dominates the excitations detected below 0.3~K. 
\end{abstract}
\maketitle

The one-dimensional $S=1/2$ quantum Heisenberg antiferromagnet (1DQHAF) described by the Hamiltonian
$
H =J \sum \boldsymbol{S}_{i}\cdot\boldsymbol{S}_{i+1}
$
continues to be of much interest, not only as a system in which quantum fluctuations
control the low-temperature physical behavior \cite{giamarchi},
but also due to the question of whether the spin transport in this system is 
ballistic or diffusive\cite{sirka,steinigeweg}. 
However, the experimental verification of theoretical work 
relies on
the existence of material systems whose interactions are well approximated
by the assumptions of the model. 
Such systems include the inorganic compound\cite{keren,motoyama} Sr$_{2}$CuO$_{3}$, the 
organometallic coordination polymer\cite{tom} Cu(pyz)(NO$_{3}$)$_{2}$ and the organic radical-ion salt\cite{francis}
DEOCC-TCNQF$_{4}$. 

Recently\cite{ishii} the inorganic system Rb$_{4}$Cu(MoO$_{4}$)$_{3}$ was identified
as another candidate quasi-one-dimensional $S=1/2$ magnetic system, where the
principle magnetic exchange occurs along one-dimensional (1D) chains, with exchange
strength $J$. Real systems do not comprise completely isolated chains
and there are also interchain interactions described by an effective exchange coupling $J_{\perp}$. 
Such a coupling, although weak in comparison to $J$ in many materials, controls the
thermodynamics and leads to long-range magnetic order (LRO)
at a nonzero temperature $T_{\mathrm{N}}$. The ratio $T_{\mathrm{N}}/J$ is therefore a good indicator
of the degree of isolation of a low-dimensional magnetic system. 
In fact, quantum Monte Carlo simulations of this effect provide a means of estimating
the effective interchain coupling $J_{\perp}$ in a quasi-1D magnet via the 
expression \cite{yasuda}
\begin{equation}
|J_{\perp}| = \frac{T_{\mathrm{N}}}{ 4c \sqrt{ \ln\left( \frac{\lambda J}{T_{\mathrm{N}}}\right)
+ \frac{1}{2}\ln \ln \left(  \frac{\lambda J }{T_{\mathrm{N}}} \right)  }  },
\label{yasj}
\end{equation}
where $c = 0.233$ and $\lambda = 2.6$. It is important to note that the 1D nature of the spin
system promotes strong thermal and quantum fluctuations which lower the magnitude of $T_{\mathrm{N}}$, 
resulting in quite long-ranged correlations building up in the chains just above the ordering temperature,
and also reduce the magnitude of the magnetic moment\cite{sengupta}. These factors often make the identification of magnetic ordering
difficult using thermodynamic probes. We have shown previously \cite{tom,tom2,steve2}
that muons are often
sensitive to LRO that can be very difficult to detect with conventional probes. At temperatures above $T_{\mathrm{N}}$
muon-spin relaxation ($\mu^{+}$SR) may also  be used to probe the nature of the spin excitations and their
transport. 
In this paper we investigate the magnetism in  Rb$_{4}$Cu(MoO$_{4}$)$_{3}$ using implanted muons. We search for long range
magnetic order down to 50~mK and investigate the spin excitations of the system. We argue that
the system remains disordered at all measured temperatures and the persistent mode that dominates the dynamics we
measure below 0.3~K is ballistic in character. 

\begin{figure}
\begin{center}
\epsfig{file=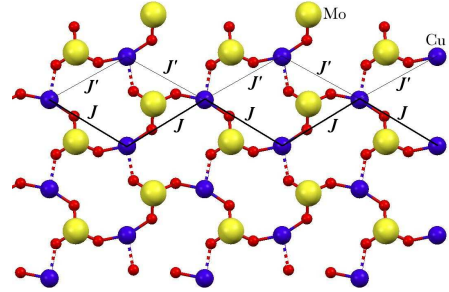,width=6cm}
\caption{The principal exchange paths in Rb$_{4}$Cu(MoO$_{4}$)$_{3}$
  (viewed along the $b$-axis) 
showing strong exchange (with strength $J$) and weak exchange
(strength $J'$) between Cu$^{2+}$ ions through Cu-O-Mo-O-Cu paths. 
Weak exchange bonds between Cu$^{2+}$ ions and apical oxygens are
dotted. 
\label{struc}}
\end{center}
\end{figure}

The structure\cite{ishii} of Rb$_{4}$Cu(MoO$_{4}$)$_{3}$ is layered, with triangular lattice sheets of $S=1/2$ Cu$^{2+}$
ions in the $a$-$c$ plane [Fig.~\ref{struc}], separated by layers of 
Rb$^{+}$ and MoO$_{4}$. Although this constrains the principal exchange
to occur along paths in the $a$-$c$ plane, the two dominant exchange
paths, with strengths $J$ and $J'$, are quite different.
The strength of  the exchange follows from
the disposition of the Cu$^{2+}$ $d_{x^{2}-y^{2}}$  orbitals\cite{ishii}. 
The local environment of the copper is a CuO$_{5}$ 
distorted square pyramid which causes the $d_{x^{2}-y^{2}}$ orbitals
to lie in the base of these pyramids. Strong $\sigma$ bonds
therefore link the $d_{x^{2}-y^{2}}$ orbitals on the Cu$^{2+}$ 
and the $p_{x,y}$ orbitals on neighbouring O$^{2-}$ ions that form the
basal plane of the pyramids, 
with one strong O-Cu-O bond in the $a$-$c$ plane and one directed along $b$. 
The main superexchange pathway, with strength $J$, 
therefore goes via two $\sigma$ bonds in the $a$-$c$
plane, along identical zig-zag Cu-O-Mo-O-Cu paths, 
as shown in Fig.~\ref{struc}. The other possible exchange 
path in this plane, which has strength $J'$, is suppressed since it
involves only one $\sigma$ bond and also 
the O-Cu bond  [shown
with dotted lines in Fig.~\ref{struc}] that involves the apical oxygen of a CuO$_{5}$ unit, whose $p_{z}$
orbitals enjoy very little overlap with the Cu$^{2+}$ $d_{x^{2}-y^{2}}$ orbital. 
In this way, the spin system appears to be formed from 
one-dimensional chains with coupling $J$, with weak interchain coupling $J'$ in the $a$-$c$ plane
and coupling between planes that is still weaker. 
This has led to the material being advanced as a strong candidate to show
 quasi-one-dimensional magnetic behavior. 
Magnetic susceptibility and heat capacity measurements\cite{ishii} confirm that the 
dominant magnetic exchange occurs along strongly interacting antiferromagnetic chains with an 
exchange constant $J=10.0$~K. No signature of LRO was detected down to
0.1~K, with only broad humps in susceptibility and heat capacity in the 4~K region, typical
of the build up of short range correlations above $T_{\mathrm{N}}$. 

\begin{figure}
\begin{center}
\epsfig{file=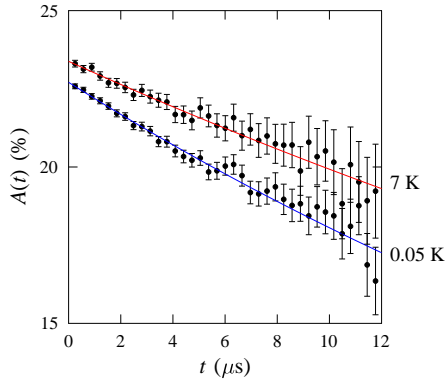,width=6cm}
\caption{ Example ZF $\mu^{+}$SR spectra measured at $T=0.05$~K
and $T=7$~K. 
\label{data}}
\end{center}
\end{figure}

In a muon-spin relaxation ($\mu^{+}$SR) experiment \cite{steve}, spin-polarized
positive muons are stopped in a target sample, where the muon usually
occupies an interstitial position in the crystal.
The observed property in the experiment is the time evolution of the
muon spin polarization, the behavior of which depends on the
local magnetic field $B$ at
the muon site, and which is proportional to the
positron asymmetry function $A(t)$. 
Our $\mu^{+}$SR 
measurements were made 
using the MuSR spectrometer at the ISIS facility, UK. 
For the measurement a polycrystalline sample of 
Rb$_{4}$Cu(MoO$_{4}$)$_{3}$
was mounted on a silver sample plate using a small amount of vacuum grease and 
the plate was attached to the cold finger of a dilution refrigerator.

\begin{figure}
\begin{center}
\epsfig{file=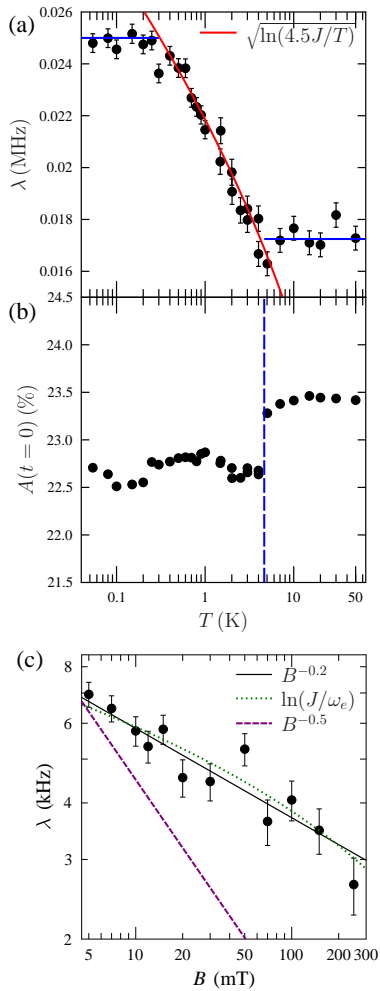,width=5.5cm}
\caption{(a) The relaxation rate $\lambda$ 
as a function of temperature is seen to increase upon
cooling through $T_{\mathrm{co}} \approx 5$~K. 
(b) The initial asymmetry $A(t=0)$ also shows a  reduction upon cooling through
$T_{\mathrm{co}}$ (dotted line).
(c) The evolution of the relaxation rate $\lambda$ with applied field (above $B=3$~mT) at $T=0.1$~K. 
All fits are described in the main text.  
\label{lambda}}
\end{center}
\end{figure}

Example zero-field (ZF) $\mu^{+}$SR spectra are shown in Fig.~\ref{data}. 
Across the entire measured temperature range ($0.05\leq T \leq 50$~K)
 $A(t)$
is found to relax exponentially with small relaxation rates
of $\lambda < 0.03$~MHz. Data were fitted to 
\begin{equation}
A(t) = A_{0}e^{-\lambda t } + A_{\mathrm{bg}},
\end{equation}
 where $A_{\mathrm{bg}}$ is a small, constant offset reflecting the contribution of those
muons that stop in the sample holder or cryostat tail. 
Oscillations, which would be
characteristic of a quasi-static local magnetic field at the 
muon stopping site, are not observed at any temperature.
However, we do observe a cross-over in magnetic behaviour in the
system upon cooling below $T_{\mathrm{co}} \approx 5$~K. 
Below this temperature the initial asymmetry drops  by an amount
$\Delta A = 0.7(1)$\% (translating to a loss of 
roughly 3\% of the total muon polarization) as shown in Fig.~\ref{lambda}(b). 
Also evident is an increase in the relaxation rate  $\lambda$, which begins to
rise below 4~K, increases steadily and then flattens off below 0.3~K [Fig.~(\ref{lambda}(a)]. 

These results are indicative of a transition or cross-over below $T_{\mathrm{co}}$
to a regime characterized by larger magnetic fields,
a broader distribution of fields or
(perhaps most likely) an increased correlation time. 
Assuming the fast-fluctuation limit (justified below) we expect\cite{hayano} the relaxation to vary as
$\lambda \propto \Delta^{2} \tau$, 
where $\Delta$  is proportional to the second moment of the field distribution via 
$\Delta^{2}= \gamma_{\mu}^{2}\langle B_{\mu}^{2}\rangle$, $\gamma_{\mu}$ is the muon gyromagnetic ratio,
$B_{\mu}$ is the magnetic field at the muon site and $\tau$ is the correlation time. 
The small fraction of asymmetry missing below $T_{\mathrm{co}}$
therefore reflects a fraction of muons which are quickly depolarized
within the first $\sim 0.1~\mu$s 
and the increased
relaxation of the remainder of the muons also indicates an increase in the magnitude or  width of their
field distribution via $\Delta$ or correlation time via $\tau$. 
The small size of the muon relaxation rate (leading to a decay time that is rather slow in 
comparison to the length of our time window) means that reliably extracting a value of the 
asymmetry baseline is difficult. We do not, however, see any evidence for an increase in the apparent
baseline on cooling through $T_{\mathrm{co}}$ as would be expected in a transition to static magnetism
in a polycrystalline sample. (Such an increase occurs since 1/3 of muon spin components would be expected to lie 
along the direction of a static local field and then not be relaxed.) This suggests that the magnetic state
is characterized by significant dynamics in the local field at the muon site(s) at all temperatures. 

To understand the nature of the muon relaxation, measurements were made in 
applied magnetic field, directed parallel to the initial muon spin polarization.
 The application of fields of $B < 3$~mT leads to a 
repolarization of the asymmetry. In view of the form of the ZF spectra, fits were carried out to a 
dynamicized Kubo Toyabe (KT) function\cite{hayano} 
with a field fluctuation rate $\nu=1/\tau$.
Fits to this function in applied field 
with a second moment of the field distribution $\Delta \approx 0.1$~MHz and a fluctuation rate 
$\nu \approx 0.7$~MHz describe the data at all measured fields below 5~mT.
The ratio $\nu/\Delta\approx 7$ places the relaxation in the fast
fluctuation limit, justifying the assumption above.

An obvious explanation for the change of behaviour around 5~K 
would be a transition to a state of LRO  at $T_{\mathrm{co}}$. 
Although one might expect that LRO would lead to oscillations in the asymmetry, 
there are several cases where such oscillations are not observed at a magnetic 
transition. Similarly the absence of an apparent increase in baseline does not preclude the occurrence of LRO either.
However, we believe LRO unlikely. 
If $T_{\mathrm{co}}=4.8(3)$~K [which we estimate from the behavior of $A(t=0)$] 
is identified with the magnetic ordering temperature $T_{\mathrm{N}}$  then 
we have a ratio $T_{\mathrm{N}}/J = 0.48(3)$. 
The model in Eq.~(\ref{yasj}) refers to a chain coupled weakly in the two perpendicular interchain directions, assumed
identical, with strength $J_{\perp}$. Although the interchain coupling in  Rb$_{4}$Cu(MoO$_{4}$)$_{3}$ is more complex,
 by applying Eq.~(\ref{yasj}) we obtain a prediction for an effective interchain coupling $|J_{\perp}| = 3.4(2)$~K
which should be expected to be of the same order of magnitude as $J'$ identified above. 
Using this estimate we find $|J_{\perp}/J| =  0.34(2)$, which  would imply a rather three-dimensional magnetic material with
little of the quite significant
exchange anisotropy suggested by the structure or the previous magnetic measurements \cite{ishii}. 
In addition, chain mean field theory gives us a predicted staggered moment \cite{schulz}
$m_{0} = 1.017 \sqrt{|J_{\perp}|/J} \approx 0.6 \mu_{\mathrm{B}}$,
which is likely to be detectable. 
It therefore seems unlikely that the weak effects we observe with muons are LRO. 
Another possibility is that small phases of impurities are present and that these are ordering. 
In this picture the small loss of asymmetry would be due to the
muons which stop in these regions and that the increased relaxation effect would
reflect the magnetic field of the impurity regions in the rest of the sample. 
This seems highly unlikely in that if such impurities are detectable with $\mu^{+}$SR 
their volume fraction must exceed several percent and their presence would likely have been detected with x-ray diffraction. 

It is most probable that the effects we observe at $T_{\mathrm{co}}$ are due to the rapid growth
of correlations in Rb$_{4}$Cu(MoO$_{4}$)$_{3}$ around 4~K.
We note that broad peaks in magnetic susceptibility and in specific heat are both observed in the 4~K region \cite{ishii}. These are
due to the build up of magnetic 
correlations within the chains and reflects the fact that the energetic 
cost of flipping a spin is
of order $J$.  This build up could conceivably lead to local
regions of well correlated spins which give rise to an increased muon relaxation. 
If that is the case then we must conclude that no LRO is observed down to 50~mK,  placing strong constraints
on the effective interchain coupling and the size of the staggered moment. 
We find 
$T_{\mathrm{N}}/J<0.005$, $|J_{\perp}| < 0.02$, $|J_{\perp}/J| < 2\times 10^{-3}$ and $m_{0}<0.05\mu_{\mathrm{B}}$. 
For comparison, we note that Cu(pyz)(NO$_{3}$)$_{2}$ has $|J_{\perp}/J|=4.4 \times 10^{-3}$, 
Sr$_{2}$CuO$_{3}$ has $|J_{\perp}/J|=0.93 \times 10^{-3}$ while
DEOCC-TCNQF$_{4}$ has $|J_{\perp}/J| < 0.06 \times 10^{-3}$. These observations imply that Rb$_{4}$Cu(MoO$_{4}$)$_{3}$
is a better example of a 1DQHAF than Cu(pyz)(NO$_{3}$)$_{2}$ and perhaps comparable to Sr$_{2}$CuO$_{3}$. 

That the physics of Rb$_{4}$Cu(MoO$_{4}$)$_{3}$ is well described by a 1D
model of spin exchange with small interchain interactions allows us to relate our results 
to the spin excitations expected for such a model. 
Below $T_{\mathrm{co}}$ [Fig~\ref{lambda}(a)] $\lambda$
increases as the temperature is reduced, and saturates
at a constant value $\lambda_{\mathrm{sat}}=0.025$~MHz below 0.3~K.
This trend of an increased relaxation rate and low-$T$ persistent dynamics 
is often seen in the $\mu^{+}$SR of complex systems
with low-temperature dynamics\cite{smms,gingras} but its origin is
not fully understood. The general behavior can arise because of the existence
of two relaxation channels, one of which is strongly $T$ dependent
with a correlation time $\tau_{\mathrm{s}}(T)$ while the other shows little
variation with $T$ and has a correlation time $\tau_{\mathrm{w}}$. 
In the presence of two competing relaxation processes
the one with the shorter correlation time wins out, giving
the smaller relaxation rate.
At higher temperature, therefore,
we have $\tau_{s}(T) \ll \tau_{w}$ which results in a strongly $T$-dependent
relaxation.  
We may gain an insight into the temperature dependent part of $\lambda$ by 
by appealing to the results of previous measurements on highly ideal 1DQHAFs. 
The muon is a local probe, so its response is due to a sum over fluctuations at
all wavevector $q$. 
However, NMR\cite{srcuo} of Sr$_{2}$CuO$_{3}$  and $\mu^{+}$SR\cite{francis} of
DEOCC-TCNQF$_{4}$ revealed that the low temperature relaxation was 
dominated by spinon excitations close to $q=\pi/a$  where low $T$ corrections to the theory of
spin correlations for the $S=1/2$ 1DQHAF result in a rise in $\lambda$ that follows $\lambda \sim \sqrt{\ln(4.5 J/T)}$.
 Such a temperature dependence provides a good description of the rapid increase in $\lambda$ at low temperatures
in Rb$_{4}$Cu(MoO$_{4}$)$_{3}$
as shown in Fig.~\ref{lambda}(a) \cite{notes}.
The increase
in relaxation rate arises from the
 lengthening of the correlation time $\tau_{s}$ with decreasing temperature. 
This increase continues until $\tau_{\mathrm{s}}$ becomes much longer than 
the correlation time $\tau_{\mathrm{w}}$
of the $T$-independent
relaxation channel with persistent dynamics, which then dominates the relaxation at the lowest temperatures,
resulting in a relaxation rate $\lambda_{\mathrm{sat}}$. 

Finally we turn to the nature of the propagation of the persistent spin
fluctuation in Rb$_{4}$Cu(MoO$_{4}$)$_{3}$. Whether the excitations in
1DQHAFs are ballistic or diffusive remains a key question
in the physics of spin chains and one that can be addressed with our LF $\mu^{+}$SR
measurements\cite{francis}. 
This is because the spectral density for ballistic motion is expected to vary with frequency $\omega$ as
$f(\omega) \sim \ln (J/\omega)$ while 1D diffusive transport results in a function
$f(\omega) \sim \omega^{-\frac{1}{2}}$.  
In applied fields $B>3$~mT at $T=0.1$~K we begin to repolarize the missing fraction of asymmetry and obtain a relaxation that
persists up to the highest applied field of $250$~mT. The result of fitting this persistent relaxation rate 
to an exponential relaxation function as a function of  applied field is shown in Fig.~\ref{lambda}(c). 
We see that 
the diffusive model, with $\lambda \propto B^{-0.5}$ provides a poor description of the 
data. In fact, fitting to an unconstrained power law form $\lambda \propto B^{n}$ results in $n=0.20(2)$, well 
below the theoretical value of $n=0.5$ and the slightly reduced value of $n=0.35$ found\cite{francis} in 
DEOCC-TCNQF$_{4}$. Turning now to the model of ballistic transport, we
note that 
the muon can couple to electronic spin density via both dipolar and
hyperfine interactions, and for ballistic spin transport we expect 
\cite{devreaux}  a relaxation rate
$
\lambda(B) = \frac{1}{20}\left[ 3 D^{2} f(\omega_{\mu}) + (5 A^{2}+7D^{2})f(\omega_{\mathrm{e}})\right],
$
where $\omega_{\mu}=\gamma_{\mu} B$, $\omega_{\mathrm{e}} = \gamma_{\mathrm{e}}B$, $\gamma_{\mathrm{e}}$
is the electron gyromagnetic ratio and $A$ and $D$ are the scalar and dipolar hyperfine coupling parameters. 
As shown in Fig.~\ref{lambda}(c), a good fit to the data is achieved by fitting 
$\lambda \propto \ln(J/\omega_{\mathrm{e}})$, which assumes that the second term in the expression for $\lambda(B)$
dominates. 
Our inference, therefore, is that the
 modes which cause the saturated relaxation at low temperature show behavior suggestive of ballistic
spin transport. 

In conclusion, our results demonstrate that Rb$_{4}$Cu(MoO$_{4}$)$_{3}$ is a good example of a quasi-1DQHAF system, with
a low ratio of exchange strengths $|J_{\perp}/J|$, 
putting it amongst the best realizations of this model. The spin
excitations are suggestive of a contribution from modes around $q=\pi/a$ in the 
region $0.3 \leq T \leq 4$~K and a persistent, 
low
temperature contribution which has a ballistic character. 

Part of this work was carried out at the STFC ISIS facility and we are grateful for the provision of beamtime. 
This work is supported by EPSRC (UK).


\begin{thebibliography}{xx}
\bibitem{notes}
Fitting this functional form to the data yields an estimate of the
exchange constant which is consistent with the $J=10.0$~K derived from
magnetic susceptibility.
\end{thebibliography}
\end{document}